\documentclass[twoside,12pt]{article}
\usepackage{amssymb}
\def\R{{\mathbb R}}

\def\N{{\mathbb N}}

\def\supp{{\mbox{\rm supp}\, }}
\def\C{{\mathbb C}}

\def\kasten{$~~\mbox{\hfil\vrule height6pt width5pt depth-1pt}$ }
\newtheorem{theorem}{Theorem}[section]

\newtheorem{corollary}[theorem]{Corollary}

\newtheorem{lemma}[theorem]{Lemma}

\begin{document}
\pagestyle{myheadings} \markboth{H. Gottschalk}{Holomorphic Random Fields} \thispagestyle{empty}
\begin{center}
 {\Large \bf Wick rotation for holomorphic random fields}

\

\noindent {\sc Hanno Gottschalk}\\
\vspace{.25cm}

{\small Institut f\"ur angewandte Mathematik, \\ Rheinische
Friedrich-Wilhelms-Universit\"at Bonn,\\ Wegelerstr. 6, D-53115
Bonn, Germany\\ gottscha@wiener.iam.uni-bonn.de}

\end{center}

{\noindent \small {\bf Abstract.} Random field with paths given as restrictions of holomorphic functions to 
Euclidean space-time can be Wick-rotated by pathwise analytic continuation. Euclidean symmetries of the correlation functions
then go over to relativistic symmetries. As a concrete example, convoluted point processes with interactions motivated from quantum field theory
are discussed. A general scheme for the construction of Euclidean invariant infinite
volume measures for systems of continuous particles with ferromagnetic interaction is given and applied to the models under consideration. Connections with Euclidean quantum field theory, 
Widom-Rowlinson and Potts models are pointed out. For the given models, pathwise analytic continuation and analytically continued correlation functions are shown to exist and to expose
relativistic symmetries. }

\

{\small \noindent {\bf Keywords}: {\it Wick rotation, Euclidean QFT, Random fields with holomorphic paths, random measures, FKG-Holley-Preston inequality, thermodynamic limit.}
 }

\section{Introduction}

\noindent {\em 1.1 Motivation}

\noindent In the very beginning of Euclidean quantum field theory (EQFT) Tadao Nakano \cite{Na} and Julian Schwinger \cite{Schw1,Schw2} proposed the analytic continuation
of time-ordered -- hence symmetric -- vacuum expectation values (VEVs) 
\begin{equation}
\tau_n(y_1,\ldots,y_n)=\left\langle 0\left|\, T\phi_{\rm rel.}(y_1)\cdots\phi_{\rm rel.}(y_n)\right| \,0\right\rangle~, ~~y_l=(y_l^0,\vec y_l)\in Y,
\end{equation}
of a relativistic, local field $\phi_{\rm rel.}(y)$ on some Lorentzian manifold $Y$ ($Y=\R^4$ in \cite{Na,Schw1,Schw2}) in all time arguments $t_l=y_l^0$ to obtain Euclidean Green's functions (Schwinger functions). No Euclidean time-ordering was used, however
different analyticity properties of time-ordered and anti-time-ordered VEVs played a r\^ole. Though this original procedure of "Wick-rotation" of (real) relativistic times to Euclidean (purely imaginary) times
was not proven in an axiomatic framework, it was the source of inspiration for later works with increasing mathematical rigorousity and generality \cite{Gl,Ne1,Ne2,OS1,OS2,Sy1,Sy2} which laid the mathematical foundations
of EQFT.      

This note takes a look at the original idea of Nakano and Schwinger from an Euclidean point of view. If one has a functional measure, representing the Euclidean field, over the Euclidean space-time $X$,
it would of course be ideal if the paths of the given field measure would all be restrictions of holomorphic functions on the complexified space-time $X^c $ to the real submanifold $X$ as one could then perform the analytic continuation
pathwisely or strongly. Analytic continuation on the level of Green's functions -- weak analytic continuation -- could then be derived by taking the ensemble average of pathwise continuations. Restricting this continuation to
relativistic space-time $Y$ embedded in $X^c $ as another real submanifold, the result would be a symmetric and relativistically invariant function which could be identified with the time-ordered VEVs of relativistic fields.

Clearly, the above described ideal situation does not apply to quantum field theory as the paths of field measures in general are not even functions \cite{CL,RR}. Nevertheless, it might be worth to look at similar -- but ultra-violet regular --
systems where the kind of Wick rotation described above can be carried through in a mathematically rigorous way. Even though such models fail to represent all features of a realistic QFT, one can still consider them as a testing ground
for phenomena which rather exclusively depend on the infra-red behavior, as e.g. spontaneous breaking of symmetries. 
Such phenomena up to now are much better understood in Euclidean QFT, where they are special cases of phase transitions of systems of statistical mechanics. The kind of Wick-rotation proposed here allows a most simple passage from Euclidean to relativistic space time and 
could therefore be useful to gain better understanding of spontaneous 
symmetry breaking on relativistic space-times. In particular, direct numerical calculation of relativistic correlation functions in principle is possible on the basis of computer simulation of the Euclidean statistical mechanics ensembles. This might allow a first glimpse,
what kind of spatial geometry one would encounter in relativistic QFT provided the related Euclidean system undergoes a breaking of rotation and / or translation symmetry.    

\

\noindent {\em 1.2 Outline}

\noindent In this work the above program is carried through in the simplest possible case, where Euclidean symmetries are unbroken and relativistic symmetries can be derived by analytic continuation. 

In Section 2 we collect some known facts about random (point) measures needed for the formulation of the Euclidean theory and about holomorphic functions in several arguments needed for the analytic continuation. 

Section 3 deals with the thermodynamic (TD) limit of interacting systems of point particles in the case where the interaction has certain "ferromagnetic" properties.
In particular, we give an existence and uniqueness result
for the infinite volume measures which automatically implies Euclidean invariance. 
This result on the TD limit probably is not new to some
experts, as similar results can also be derived from tightness arguments. However, it seems that they have not been written down in detail and so a proof is supplied
for the convenience of the reader. We introduce a class of models, where ferromagneticity can be verified.    
 
In Section 4 we show how the models of Section 3 are motivated by EQFT \cite{AGW,AGY1,AGY2}:   
Convolution of a random point measure representing a statistical ensemble of  particles with a real-valued integral
kernel $G$ takes over the system of point particles into a continuous field system. In particular, the interactions of 
 the models in Section 3 can be understood in terms of local interactions for the Euclidean field. Relations with
models of Widom-Rowlinson type and Potts models are clarified.

Section 5 considers the situation of Section 4 when the integral kernel $G$ is the restriction to $X\times X$ of a kernel 
$G^c$ defined on $X^c \times X$ which is complex analytic in the first argument. If $G^c$ is of sufficiently fast decrease in the second argument,
one can show that the paths of the obtained Euclidean field have analytic extension from $X$ to $X^c $. It is then not difficult to prove that also the
$n$-point moment (Schwinger) functions $S_n$ of the Euclidean random fields have analytic continuation $S_n^c$ from $X^n$ to $(X^c)^n$. 
If in addition invariance of the kernel $G^c$ under Euclidean transformations given by some real Lie group ${\cal G}$ acting on $X$ with complexification ${\cal G}^c$ on $X^c $ is assumed, one can apply arguments from axiomatic quantum field theory \cite{SW,V} to prove that the analytically continued Schwinger functions $S^c_n$ are invariant under ${\cal G}^c$. Here Euclidean invariance of
the infinite volume measure of the particle systems enters. Restricting
$S_n^c$ from $(X^c) ^n$ to $Y^n$ and denoting this restriction by $\tau_n$ where $Y\subseteq X^c $ is the "relativistic" real submanifold, one obtains that the Wick-rotated functions $\tau_n$ are invariant under the
subgroup ${\cal G}^r$ of ${\cal G}^c$ stabilizing $Y$ as a set. 
We give concrete examples on flat Minkowski or expanding de Sitter space-times 
in order to identify ${\cal G}^r$ with the known groups of relativistic symmetries. A short discussion on antiholomorphic extensions and the potential failure of relativistic invariance of the Wick-rotated functional measure
on the space of complex valued functions on $Y$ concludes the article.

\section{Preparations}
In this part we fix our notations and collect some known results from the theory of random (point) measures and holomorphic functions.

\

\noindent{\em 2.1 Random point measures}

Let $X$ be a locally compact topological second countable space equipped with the Borel-$\sigma$-algebra ${\cal B}(X)$,
and let $C_b(X)$ be the space of bounded, continuous, real valued functions on $X$ and $C_0^+(X)$ the subspace of non-negative, compactly supported continuous functions.  By ${\cal M}={\cal M}(X)$ we denote 
the space of (non-negative) Radon measures on $X$ equipped with the
Borel-$\sigma$-algebra ${\cal B}({\cal M})$ generated by the vague topology, thus the $\sigma$-algebra geberated by $N_h:\eta\to\langle\eta,f\rangle=\int_Xh\,d\eta$ 
$h\in C_0^+(X)$. ${\cal M}$ is partially  ordered by writing $\eta\leq \gamma$ if $\eta(A)\leq\gamma(A)$ $\forall A\in {\cal B}(X)$. 

A Laplace transform is a positive-definite, continuous and normalized function from $C_0^+(X)$ to $[0,1]$. By \cite[Proposition I.12]{Ne} there is a one-to-one correspondance 
between the Laplace transforms ${\cal L}$ on $C_0^+(X)$ and random measures, i.e. probability measures $\mu$ on $({\cal M},{\cal B}({\cal M}))$, given by
\begin{equation}
\label{2.1eqa}
{\cal L}(h)=\int_{\cal M}e^{-\langle \eta,h\rangle}\, d\mu(\eta)\, , ~~h\in C_0^+(X).
\end{equation} 
By definition, a sequence of random measures $\mu_n$ converges in distribution to the random measure $\mu$ if $\int_{\cal M}F\,d\mu_n\to\int_{\cal M}F\, d\mu$ $\forall F:{\cal M}\to\R$
bounded and vaguely continuous. The following establishes a connection between the convergence of Laplace transforms and convergence in distribution:

\begin{theorem}
\label{2.1theo}
Let $\mu_n$ be a sequence of random measures and ${\cal L}_n$ the associated sequence of Laplace transforms. If for $h\in C_0^+(X)$ the limit ${\cal L}_n(h)\to{\cal L}(h)$ exists and the function
$[0,\infty)\ni t\to{\cal L}(th)$ is continuous at zero then there exists a random measure $\mu$ such that ${\cal L}$ is the Laplace transform of $\mu$ and $\mu_n\to\mu$ in distribution.
\end{theorem}
For the proof see \cite[Corollaire I.17]{Ne}.

The subset of point measures $\Gamma\subseteq {\cal M}$ is the space of positive measures $\eta$ s.t. 
$\eta(A)\in \bar \N_0=\N_0\cup\{\infty\}$ $\forall\, A \in{\cal B}(X)$ and $\eta(A)<\infty$ if $A$ is relatively compact in $X$.
$\Gamma$ is equipped with the $\sigma$-algebra  ${\cal B}(\Gamma)$ generated by the counting variables $N_A:\eta\to\eta(A)$, $A\in{\cal B}(X)$.
It is well-known that ${\cal B}(\Gamma)$ is the trace-$\sigma$-algebra of $\Gamma$ in $({\cal M},{\cal B}(M))$,  cf. \cite[Lemma 4.1]{Ka}. A random point measure
by definition is a probability measure on $({\cal M},{\cal B}({\cal M}))$ with support contained in $\Gamma$. Equivalently, it can be defined as a probability measure
on $(\Gamma,{\cal B}(\Gamma))$. One gets from \cite[Chapter 4, p. 21]{Ka}:

\begin{theorem}
\label{2.2theo}
 If sequence of random point measures $\mu_n$ converges in law to a random measure $\mu$, then the limiting measure $\mu$ also is a random point measure. 
\end{theorem}

 For $\lambda$ a Radon measure on $(X,{\cal B}(X))$, 
let $\mu_\lambda$ be the Poisson measure with intensity measure $\lambda$ on $(\Gamma,{\cal B}(\Gamma))$. This process is uniquely
 determined by its Laplace transform ${\cal L}(f)=\exp\{\int_X(1-e^{-f})\, d\lambda\}$. 
 If another such measure $\sigma$ is even globally finite, then $\mu_\sigma$
has support on $\Gamma_0$, the set of finite point measures over $X$. ${\cal B}(\Gamma_0)$ is the trace sigma algebra of ${\cal B}(\Gamma)$ on $\Gamma_0$.  

 A function $F:{\cal M}\to\R$ ($F:\Gamma\to\R$) is called increasing, if $F(\eta)\leq F(\gamma)$ for $\eta\leq \gamma$, $\eta,\gamma\in\Gamma$ ($\eta,\gamma\in{\cal M}$). A set $A\in {\cal B}({\cal M})$ ($A\in{\cal B}(\Gamma)$) is called increasing, if its characteristic function is increasing.
A random (point) measures $\mu_2$ dominates another such measure $\mu_1$ in stochastic order, in symbols $\mu_1\preceq\mu_2$, if $\int_{\cal M}Fd\mu_1\leq\int_{\cal M}Fd\mu_2$ ($\int_{\Gamma}Fd\mu_1\leq\int_{\Gamma}Fd\mu_2$) for all $F:{\cal M}\to\R$ ($F:\Gamma\to\R$) measurable, bounded and increasing. 
If this inequality holds for $F$ increasing and vaguely continuous, but not necessarily for all increasing measurable $F$, $\mu_1$ is vaguely dominated in stochastic order by $\mu_2$, in symbols
$\mu_1\prec_v\mu_2$. 

Following \cite{GK}, the class ${\cal P}_\sigma$ of random point measures is introduced as those random point measures $\mu$ that are absolutely continuous w.r.t. $\mu_\sigma$ such that
the Radon-Nikodym derivative $f$ admits a version such that the set $\{f=0\}$ is increasing. Let an arbitrary such version be fixed. A version of the Papangelou intensity of $\mu$ w.r.t. to $\mu_\sigma$ is defined a measurable function 
$p:X\times \Gamma\to \R$ such that  $p(x,\eta)f(\eta)=f(\delta_x+\eta)$. Here $\delta_x$ is the Dirac measure with mass one in $x$. The main result of \cite{GK} is the FKG-Holley-Preston inequality \cite{FKG,H,P} for continuous particle systems:

\begin{theorem}
\label{2.3theo}
(i) Let $\sigma\in{\cal M}$ be finite and $\mu_1,\mu_2\in{\cal P}_\sigma$ admit versions of Papangelou intensities $p_1,p_2$ such that
$p_1(x,\eta)\leq p_2(x,\gamma)$ for all $\eta,\gamma\in\Gamma$, $\eta\leq \gamma$ and $x\in X\setminus \supp(\gamma-\eta)$. Then $\mu_1\preceq\mu_2$.

\noindent (ii) In particular, if $\mu\in {\cal P}_\sigma$ admits a Papangelou intensity $p$ such that $p(x,\eta)\leq\rho(x)$ $\forall\eta\in\Gamma,x\in X$ with $\rho\in L^1(X,\sigma)$ then
$\mu\preceq\mu_{\rho\sigma}$. 
\end{theorem}

\noindent{\em 2.2 Holomorphic functions}

\noindent For the convenience of the reader, some classical theorems on analytic functions are recalled. 

\begin{theorem}[\cite{GR}]
\label{2.4theo}
Let $\Omega\subseteq \C^n$ a domain and $g_n:\Omega\to\C$ a sequence of holomorphic functions with $\lim_{n\to\infty}g_n(z)=g(z)$ $\forall z\in\Omega$ where
the limit is uniform on compact subsets in $\Omega$. Then $f:\Omega\to\C$ is holomorphic. 
\end{theorem}

The following is an obvious generalization of Morera's theorem to the case of several complex variables. It easily follows from
Morera's theorem in one variable, the continuity Lemma \cite[Lemma 16.1]{ER} and the fact that a function $f:\Omega\to\C$ is holomorphic iff
$f$ is holomorphic in each argument separately and jointly continuous in all arguments (Theorem of Osgood \cite{GR}).  

\begin{theorem}
\label{2.5theo}
Let $\mu$ be a finite measure on a measurable space $(\Sigma,{\cal B})$ and $g:\Omega\times \Sigma\to\C$, with $\Omega\subseteq \C^n$ a domain such that 
(i) $\Sigma\ni \eta\to g(z,\eta)$ is measurable $\forall z$; (ii) $\forall z$ there exists an open ball $B\subseteq \Omega$ containing
$z$ and a function $H\in L^1(\Gamma,{\cal B}(\Gamma),\mu)$ such that $|g(z,\eta)|\leq H(\eta)$ $\forall \eta\in \Sigma, z\in B$; (iii) $g(.,\eta):\Omega\to\C$ is holomorphic
for $\mu$ a.e. $\eta\in\Gamma$. Then $\tilde g(z)=\int_\Sigma g(z,\eta)d\mu(\eta)$ defines a holomorphic function $\tilde g:\Omega\to\C$.   
\end{theorem}

By use of local charts it is clear that Theorems \ref{2.4theo} and \ref{2.5theo} equally hold for $\Omega$ a domain
in a complex analytic  manifold. 

To prove relativistic invariance in Section 5, one also needs the following result about the extension of invariance under the action of a real Lie group to the complexified Lie group:
\begin{theorem}
\label{2.6theo}
Let ${\cal G}$ be a real analytic Lie group and let ${\cal G}^c$ its complexification. If $f:{\cal G}^c\to \C$ is holomorphic and
$f$ is constant on ${\cal G}$, then $f$ is constant on ${\cal G}^c$.
\end{theorem}

The proof of Theorem \ref{2.6theo} is given in \cite[Lemma 4.11.13]{V}.

\section{Ferromagneticity and TD limit}

\noindent {\em 3.1 A general strategy}

\noindent Let $f:\Gamma_0\to(0,\infty)$ be a ${\cal B}(\Gamma_0)$-measurable function with Papangelou
 intensity $p(x,\eta)=f(\delta_x+ \eta)/f(\eta)$, $\eta\in\Gamma_0$, $x\in X$, such that the following conditions hold
\begin{itemize}
\item[(i)] {\em Stability:} $f(\eta)\leq \xi^{\sharp \eta}$ $\forall \eta\in \Gamma_0$ for some $\xi >0$;

\item[(ii)] {\em Stochastic upper bound:} $\exists$ a positive function $\rho\in L^1_{\rm loc.}(\R^d,\lambda)$ such that $p(x,\eta)\leq \rho(x)$ for all $\eta\in\Gamma_0$, $x\in X$;

\item[(iii)] {\em Ferromagneticity:} $p(x,\eta)\leq p(x,\gamma)$ for $\eta,\gamma\in\Gamma_0$, 
$\eta\leq\gamma$, $x\in X\setminus\supp(\gamma-\eta)$.
\end{itemize}

In the limit $\sigma\nearrow\lambda$, $\sigma$ is assumed to be a finite measure on $(X,{\cal B}(X))$ s.t.
$\sigma\leq\lambda$. By (i), $f\in L^1(\Gamma_0,d\mu_\sigma)$ and one can define $d\mu_\sigma^f=fd\mu_\sigma/\int_{\Gamma_0}fd\mu_\sigma$. 

In this subsection the following result is proven:

\begin{theorem}
\label{3.1theo}
If $f$ fulfills conditions (i)---(iii), then there exists a uniquely determined measure $\mu_\lambda^f$ on $(\Gamma,{\cal B}(\Gamma))$ such that $\mu_\sigma^f\to\mu_\lambda^f$ weakly as $\sigma\nearrow\lambda$. 
\end{theorem}

The proof of the theorem is started with the following lemma: 

\begin{lemma}
\label{3.1lem}
Let $\sigma_j$ fulfill the conditions given on $\sigma$ above for $j=1,2$. Then $\sigma_1\leq\sigma_2$ implies $\mu^f_{\sigma_1}\preceq\mu_{\sigma_2}^f$.
\end{lemma}
\noindent {\bf Proof\footnote{The argument, short-cutting a previous proof, is due to H. O. Georgii.}.} Note that $\mu_{\sigma_j}^f\in{\cal P}_{\sigma_2}$ for $j=1,2$. The Papangelou intensity of $\mu_{\sigma_1}^f$ w.r.t. $\mu_{\sigma_2}$
admits a version $p_1(x,\eta)=p(x,\eta)(d\sigma_1/d\sigma_2)(x)$ with the Radon-Nikodym derivative $(d\sigma_1/d\sigma_2)\leq 1$. $\mu_{\sigma_2}^f$ admids Papangelou intensity $p_2(x,\eta)=p(x,\eta)$.
By condition (iii) this implies $p_1(x,\eta)\leq p_2(x,\gamma)$ $\forall \eta,\gamma\in\Gamma$, $\eta\leq\gamma$, $x\in X$. Now application of Theorem \ref{2.3theo} (i) completes the proof. \kasten

\noindent {\bf Proof of Theorem \ref{3.1theo}:} Let $F(\eta)=\exp\{-\langle \eta,h\rangle\}$, $h\in C_0^+(X)$. Then $-F$ is bounded, measurable and increasing. Thus, by Lemma \ref{3.1lem}, $0\leq{\cal L}_{\sigma_2}^f(h)\leq{\cal L}_{\sigma_1}^f(h)$ if $\sigma_1\leq\sigma_2$ with ${\cal L}_\sigma^f$ the 
Laplace transform of $\mu_\sigma^f$. The limit
${\cal L}_\lambda^f(h)=\lim_{\sigma\nearrow\lambda}{\cal L}_\sigma^f(h)$ thus exists by monotonicity $\forall h\in C_0^+(X)$.

By the stochastic upper bound condition (ii) and Theorem \ref{2.3theo} (ii) we obtain
\begin{eqnarray}
\label{3.1eqa}
1-L_\sigma^f(th)&\leq& t\int_\Gamma\langle\eta,h\rangle\, d\mu_\sigma^f(\eta)\nonumber\\
&\leq& t \int_\Gamma \langle\eta,h\rangle \,d\mu_{\rho\lambda}=t\int_X h\,\rho d\lambda.
\end{eqnarray}
As this estimate is uniform in $\sigma$ for $\sigma\leq \lambda$, one gets ${\cal L}_\lambda^f(th)\to 1$ for $t\searrow 0$ and $h\in C_0^+(X)$ arbitrary.
By application of Theorems \ref{2.1theo} and \ref{2.2theo} we now obtain the weak convergence of $\mu_\sigma$ as $\sigma\nearrow\lambda$ to the uniquely determined random point measure
$\mu_\lambda^f$ with Laplace transform ${\cal L}_\lambda^f$.\kasten  

It has been known for a long time \cite[Lemma 4.5]{Ka} that tightness of a sequence $\mu_{\sigma_n}^f$, $\sigma_n\nearrow \lambda$, is equivalent to 
\begin{equation}
\label{3.2eqa}
\lim_{t\to\infty}\limsup_{n\to\infty} \int_\Gamma 1_{\{\eta(A)>t\}}(\eta)\, d\mu_{\sigma_n}^f(\eta)=0
\end{equation}
for all bounded $A\in{\cal B}(X)$. As $ \int_\Gamma 1_{\{\eta(A)>t\}}(\eta)\, d\mu_{\sigma_n}^f(\eta)\leq (1/t)\int_\Gamma \eta(A)\, d\mu_{\rho\lambda}=(1/t)\int_A \rho d\lambda$ by (ii) and Theorem \ref{2.3theo} (ii), it is clear that the
stochastic upper bound alone suffices to prove the existence of weak limit points. Ferromagneticity (iii) then accounts for uniqueness of the limit. 
 
Let $\alpha:X\to X$ be a diffeomorphism with canonical action $T_\alpha$ on $\Gamma$. Note that $T_\alpha^*d\mu_\sigma=d\mu_{T_\alpha\sigma}$. $f$ is $\alpha$-invariant
if $f(\eta)=f(T_\alpha\eta)$.
As an immediate corollary the uniqueness statement of Theorem \ref{3.1theo} and the equivalence of $\sigma\nearrow \lambda$ and $T_\alpha\sigma\nearrow \lambda$ for $\lambda$ $\alpha$-invariant one gets:
\begin{corollary}
\label{3.1cor}
Let $\alpha:X\to X$ be a continuous bijection leaving $\lambda$ and $f$ $\alpha$-invariant. Then $\mu_\lambda^f$ is invariant under $\alpha$, i.e. $\mu_\lambda^f=T_\alpha^*\mu_\lambda^f$.
\end{corollary}

Another remark concerns the properties of the limiting measure:

\begin{corollary}
\label{3.2cor}
$\mu_\lambda^f$ is vaguely dominated by $\mu_{\rho\lambda}$ in stochastic order and fulfills the FKG corellation inequality 
\begin{equation}
\label{3.3eqa}
\int_\Gamma \prod_{j=1}^nF_j\, d\mu_\lambda^f\geq\prod_{j=1}^n\int_\Gamma F_j\, d\mu_\lambda^f
\end{equation}
for $F_j$ bounded, vaguely continuous, increasing and ${\cal B}(\Gamma)$ measurable.
\end{corollary} 
\noindent{\bf Proof.} From (iii) and Theorem \ref{2.3theo}  it follows that (\ref{3.3eqa}) holds for $\mu_\lambda^f$ replaced with $\mu_\sigma^f$, $\sigma\leq\lambda$ finite, cf. \cite[Corollary 1.2]{GK}. Convergence in law, Theorem \ref{3.1theo}, implies that 
both sides converge as $\sigma\nearrow\lambda$.\kasten

\

\noindent {\em 3.2 Weakly attractive interaction} 

\noindent It is not difficult to verify the conditions (i)--(iii) for a number of examples. In the following $f=e^{-\beta U}$ for some potential $U:\Gamma_0\to\R$ and
some inverse temperature $\beta>0$.
  
Let $G:X\times X\to[0,\infty)$ be symmetric, $G(x,y)=G(y,x)$, $G(.,x)\in L^1(X,\lambda)$ for $x\in X$ such that 
 $\sup_{x\in X}\|G(.,x)\|_{L^1(X,\lambda)}\leq C$ for some $0<C<\infty$ and $v:[0,\infty)\to\R$ measurable, concave and linearly bounded, i.e. $|v(\phi)|\leq b|\phi|$ $\forall \phi\in[0,\infty)$ and some $0<b<\infty$.
 Setting $G*\eta(x)=\int_X G(x,y)d\eta(y)$ one defines the potential energy of the model
\begin{equation}
\label{4.1eqa}
U(\eta)=\int_{X}v(\phi)\, d\lambda\, ,~~\phi=G*\eta, ~~\eta\in\Gamma_0.
\end{equation}
Obviously, $|U(\eta)|\leq B\sharp\eta$, $B=bC$, which gives (i) for $\xi=e^{\beta B}$. Furthermore, $\gamma(x,\eta)=e^{-\beta [W(x,\eta)+U(\delta_x)]}$ where $W(x,\eta)=U(\delta_x+\eta)-U(\delta_x)-U(\eta)$. In the present 
case the interaction $|W(x,\eta)+U(\delta_x)|\leq B$, hence (ii) holds for $\rho(x)=e^{\beta B}$ constant. (iii) is equivalent with $W(x,\eta)+U(\delta_x)$ being monotonically decreasing in $\eta$. Let $\eta\leq\gamma$,
then $\Phi_1=G*\eta\leq G*\gamma=\Phi_2$ on $X$. With $\phi=G*\delta_x$, one gets from $v$ being concave
\begin{eqnarray}
\label{4.2eqa}
W(x,\eta)+U(\delta_x)&=&\int_{X} [v(\phi+\Phi_1)-v(\Phi_1)]\,d\lambda\nonumber\\
&\geq& \int_{X} [v(\phi+\Phi_2)-v(\Phi_2)]\,d\lambda=W(x,\gamma)+U(\delta_x) \, .
\end{eqnarray}
Hence (iii) holds. If $X$ is a Riemannian manifold $\lambda$ a positive multiple of the canonic volume form and $G$ invariant
under isometries $\alpha$, i.e. $G(\alpha(x),\alpha(y))=G(x,y)$, $x,y\in X$, then $U$ is invariant under the isometry group of 
$X$ and by Corollary \ref{3.1cor} the same applies to $\mu_\lambda^f$.

Note that for the special choice $v(\phi)=1-e^{-\phi}$ one obtains interaction of Widom-Rowlinson type \cite{GK}. The uniqueness statement of Theorem \ref{3.1theo} is not in contradiction
with the observed phase transition in \cite{GH} as in Theorem \ref{3.1theo} one has specific (empty) boundary conditions.

The weakly attractive interaction has been introduced in \cite{Go} for the case of particles carrying a random charge, which goes beyond the method\footnote{If one considers the charged or "marked" random process
as a point process on the product of $X$ and the mark space one could still try to apply \cite{GK}, but this leads to a different notion of stochastic comparison as this would neglect the natural ordering on the mark space. 
In particular, it would fail to produce ferromagneticity in the case where
charges with opposite signs occur \cite{Go}.} of \cite{GK}.  

\section{Connection with field theory}

\noindent Before we carry on towards analytic continuation, we add some remarks on the physical motivation for the weakly attractive interaction of Subsection 3.2.

\

\noindent {\em 4.1 Convoluted random point measures and local interactions}

\noindent The convolution $\eta\to \phi d\lambda=(G*\eta)d\lambda$ then induces a measurable mapping from $(\Gamma_0,{\cal B}(\Gamma_0))$ to $({\cal M},{\cal B}({\cal M}))$. 
	   The image measure of a point measure $\mu$ with support on $\Gamma_0$ under this mapping is 
a random measure $\nu$. In certain situations described in \cite{AGY1,AGY2} such a random measure $\nu$ can be identified with a Euclidean quantum field of Poisson type. In fact, if  $X=\R^d$ and $\lambda$ the Lebesgue measure, $\mu=\mu_\lambda$ is pure Poisson and $G$ is a special
invariant Bessel function,
the moments $\nu=\nu_\lambda$ can be analytically continued to relativistic Wightman functions on Minkowski space \cite{AGW} (the case of de Sitter space has also been studied, see \cite{GT}).

If we take $\mu=\mu_\sigma^f$ with $f$ defined in Subsection 3.2, one obtains for $\nu=\nu_\sigma^f$
\begin{equation}
\label{5.1eqa}
d\nu_\sigma^f(\phi)={\exp\{-\beta \int_X v(\phi)\, d\lambda\}\over \int_{\cal M}\exp\{-\beta \int_X v(\varphi)\, d\lambda\}\, d\nu_\sigma(\varphi)}\, d\nu_\sigma(\phi)
\end{equation} 
Here $\nu_\sigma$ is the image measure of $\mu_\sigma$ under $\eta\to\phi=G*\eta$ and we identify $\phi d\lambda$ with the density function $\phi:X\to[0,\infty)$. 
The interaction now obviously is a local action term in (classical or quantum) field theory. 

Having found a solution for the infinite volume limit of the system $\mu_\sigma^f$ with $f$ specified in Section 3.2, one immediately gets:
\begin{corollary}
\label{5.1cor}
The family of random measures $\nu_\sigma^f$, $\sigma\leq \lambda$ finite, converges weakly to a uniquely determined random measure
$\nu_\lambda^f$ as $\sigma\nearrow\lambda$.
\end{corollary}
{\bf Proof.} Note that by definition of the image measure and symmetry of $G$, the Laplace transform of $\nu_\sigma^f$ evaluated on $h\in C_0^+(X)$ is ${\cal L}_\sigma^f(G*h)$. It is easily seen the argument
in the proof of Theorem \ref{2.1theo} extends to non-negative functions $h\in L^1(X,\rho \lambda)$ where in the case under consideration $\rho$ is a constant, cf. Section 3.2. 
By the conditions on $G$, $C_0^+(X)\ni h\to G*h\in L^1(X,\lambda)$ is continuous,
thus it follows that the limit of the Laplace transform of $\nu_\sigma^f$ exists and is continuous at zero. Theorem \ref{2.1theo} then gives the convergence statement.  \kasten
   
Theorem \ref{3.1theo} in combination with Section 4.1 is thus of interest for the solution of the infra-red problem for certain Euclidean quantum field theories of Poisson type. Furthermore, stochastic comparison and ferromagneticity survive
the passage from random point measures to field measures:

\begin{corollary}
\label{5.2cor}
(i) Let $\mu_1\preceq\mu_2$ be two stochastically ordered point random fields and $\nu_1,\nu_2$ the image field measures. Then $\nu_1\preceq\nu_2$.

\noindent (ii) If $\mu$ is a point random field which fulfills the FKG-inequality (\ref{3.3eqa}), then also $\nu$ fulfills this inequality.

\noindent (iii) In particular, the infinite volume field measure $\nu_\lambda^f$ with $f$ as in Section 4.1 fulfills the FKG inequality. 
\end{corollary}
\noindent {\bf Proof.} The proof is an immediate consequence from the fact that for $F:{\cal M}\to\R$ increasing, $F^G:\Gamma\to\R$ given by
$F^G(\eta)=F(G*\eta)$ is increasing. \kasten   

\

\noindent {\em 4.2 Connections to Widom-Rowlinson and Potts models}

\noindent It might be worth noting that interactions of Widom-Rowlinson type quite naturally occur in Poisson Euclidean QFT: To clearify this, let $\mu_{\lambda,r}$ be the random measure
with Laplace transform ${\cal L}(h)=\exp\{\int_X \omega(h)\, d\lambda\}$ with $\omega(h)=\int_0^\infty (1-e^{- s h})dr(s)$ where $r$ is a probability measure on $(0,\infty)$. $\mu_{\lambda,r}$ is thus a
marked Poisson process, where in each point of the Poisson process a random charge distributed according to $r$ is fixed. The measures on which $\mu_{\sigma,r}$ is supported thus are of the form
$\gamma=\sum_{j=1}^\infty s_j\delta_{x_j}$ with $s_j\in \supp r\cup \{0\}$ and $\{x_j\}$ a sequence of points in $X$ without accumulation points. We define $\nu_{\lambda,r}$ as the image measure 
under convolution with a kernel $G$ and $\nu_{\sigma}$ as above. $\nu_{\lambda,r}$ and $\nu_\sigma$ can be considered as free field measures. With the simplest possible local coupling of these field systems
one gets as an interacting measure
\begin{equation}
\label{5.2eqa}
d\nu(\phi,\varphi)={\exp\{-\beta\int_X\phi\varphi\, d\lambda\}\over \int_{{\cal M}\times{\cal M}}\exp\{-\beta\int_X\phi'\varphi'\, d\lambda\}\,d(\nu_\sigma\otimes\nu_{\lambda,r})(\phi',\varphi')}\, d(\nu_\sigma\otimes\nu_{\lambda,r})(\phi,\varphi)
\end{equation}
Projection on the fist component (integrating (\ref{5.2eqa}) over $\varphi$) yields $\nu_\sigma^f$, cf. (\ref{5.1eqa}), with $v(\phi)=\omega(\beta\phi)$. This interaction again fulfills the requirements of Subsection 4.1.  

If in this construction we choose $\beta=1$ and $r=\delta_1$, $v$ coincides
with the density function for the Widom-Rowlinson model. That the projection of the measure $\nu$ onto its
first component leads to such a kind of "effective action" is however not surprising. It is well-known, see e.g. \cite{GH}, that
the Widom-Rowlinson type of models occur as projection to one particle species of a continuum Potts model. Looking at the
field theoretical action $\int_X\phi\varphi\, d\lambda$ on the level of particle systems, $\phi=G*\eta$, $\eta=\sum_{j=1}^N\delta_{y_j}$, $\varphi=G*\gamma$, $\gamma=\sum_{j=1}^\infty s_j\delta_{x_j}$ one obtains
\begin{equation}
\label{5.3eqa}
U(\eta,\gamma)=\int_X\phi\varphi\, d\lambda=\sum_{j=1}^N\sum_{k=1}^\infty s_kG_1(y_j,x_k)
\end{equation}  
where $G_1(x,y)=G*G(x,y)=\int_X G(x,z)G(z,y)d\lambda(z)$. $U(\eta,\gamma)$ therefore describes the interaction of a continuum Potts model where
two species of particles (one of them "marked") interact through mutual pair-repulsion, but do not self-interact.

\section{Holomorphic paths and Wick rotation}

\noindent {\em 5.1 Pathwise analytic continuation}

\noindent Let $X^c$ be a second countable complex manifold such that $X\subseteq X^c$ is embedded into $X^c$ as a real manifold. Let $G^c:X^c\times X\to \C$ have the following
properties:

\begin{itemize}
\item[(I)] $G=G^c|_{X\times X}$ fulfills the requirements of Section 3.2;
\item[(II)] $\forall K\subseteq X^c$ compact $g_K(x)=\sup_{z\in K}|G^c(z,x)|$ is in $ L^1(X,\lambda)\cap C_b(X)$;
\item[(III)] $G^c(z,x)$ is measurable in $x$ for $z$ fixed and is holomorphic in $z$ for $x$ fixed. 
\end{itemize}

The above assumptions imply:

\begin{theorem}
\label{6.1theo}
$\mu_\lambda^f$ defined as in Section 4.1 has a modification with holomorphic paths, i.e. there exists a subset ${\cal H}\subseteq {\cal M}$ with $\mu_\lambda^f$-outer measure equal to one such that
for $m\in{\cal H}$ we have $m=\phi d\lambda$, where $\phi:X\to[0,\infty)$ is the restriction of a holomprphic function $\phi^c:X^c\to\C$ to $X$.
\end{theorem}

\noindent {\bf Proof.} Let $g\in L^1(X,\lambda)\cap C_b(X)$. Then, by the vague stochastic upper bound, cf. Corollary \ref{3.2cor},
\begin{equation}
\label{6.1eqa}
\int_\Gamma \langle\eta,g\rangle \, d\mu_\lambda^f(\eta)\leq  \int_\Gamma \langle\eta,g\rangle \, d\mu_{\rho \lambda}(\eta)=\rho \int_X g \, d\lambda <\infty
\end{equation}
with $\rho=e^{\beta B}$ as in Section 3.2. To see this in detail, let first $\langle\eta,g\rangle$ in the first inequality be replaced by $F=\max\{\langle \eta,\tilde g\rangle,M\}$ for $M>0$ arbitrary and $\tilde g\in C_0^+(X)$.
Then $F$ is continuous, increasing and vague-continuous and the inequality thus holds. By monotone convergence, one can then take in the first inequality in (\ref{6.1eqa}) the limit $M\to\infty$ and then extend it to all non-negative $g\in L^1(X,\lambda)\cap C_b(X)$. 
Thus, for every such $g$, $\Gamma_g=\{ \eta\in \Gamma: \langle\eta,g\rangle<\infty\}$ is measurable and $\mu_\lambda^f(\Gamma_g)=1$.

Next it is proven that there exists a measurable subset $\Pi\subseteq \Gamma$ such that for all $K\subseteq X^c$ compact and $\eta \in \Pi$ one gets $\langle\eta,g_K\rangle<\infty$ with $g_K$ as in condition (II). 
In fact, let $K_n\subseteq X^c$ be compact subsets of $X^c$ with $\cup_{n\in\N}K_n=X^c$, $\sharp\{n\in\N:K_n\cap K\not=\emptyset\}<\infty$ for all compacts $K\subseteq X^c$. 
Let $\Pi=\cap_{n\in\N} \Gamma_{g_{K_n}}$ then $\Pi$ is measurable and $\mu_\lambda^f(\Pi)=1$. For $K\subseteq X^c$ an arbitrary compact subset and $\eta\in \Pi$ one gets $\langle \eta,g_K\rangle\leq 
\sum_{K\cap K_n\not=\emptyset} \langle \eta,g_{K_n}\rangle<\infty$.    

Note that $\tilde G:\Pi\ni\eta\to m=(G*\eta)d\lambda\in{\cal M}$ is well-defined and measurable. This mapping can be extended to $\Gamma$ by setting the result equal to $0$ for $\eta\not\in\Pi$. 
Let ${\cal H}=\tilde G(\Pi)$. For any $A\in{\cal B}({\cal M})$ with ${\cal H}\subseteq A$ we get $\Pi\subseteq \tilde G^{-1}(A)$ and thus $\nu_\lambda^f(A)=\mu_\lambda^f(\tilde G^{-1}(A))=1$, as $\nu_\lambda^f$
is the image measure of $\mu_\lambda^f$ under $\tilde G$. The latter fact can be seen from the coincidence of the Laplace transforms, cf. the proof of Corollary \ref{5.1cor}. 
Consequently, ${\cal H}$ is of $\nu_\lambda^f$-outer measure one.

It remains to prove that for $\eta \in \Pi$, $\phi=G*\eta$ has a holomorphic extension from $X$ to $X^c$. To this aim we set $\phi^c(z)=G^c*\eta(z)=\int_XG(z,x)d\eta(x)$ which is well-defined as for $z\in K$, $K$ a compact subset in $X^c$, we have
$|G(z,x)|\leq g_K(x)$ $\forall x\in X$ and $g_K\in L^1(X,\eta)$. It remains to show that $\phi^c$ is holomorphic on $X^c$. Let $\Lambda_n\subseteq X$ a monotonically increasing sequence of bounded sets s.t. $\Lambda_n\nearrow X$ and
$\eta_n=1_{\Lambda_n} \eta$ with $1_A$ the characteristic function of $A\subseteq X$. Clearly, $\phi^c_n=G^c*\eta_n$ is holomorphic on $X^c$ as a finite sum of holomorphic functions, see (III). By Theorem  \ref{2.4theo} the holomorphy of $\phi^c(z)$
follows from $\sup_{z\in K}|\phi(z)-\phi_n(z)|\to 0$ as $n\to\infty$. To see this, one observes that
\begin{equation}
\label{6.2eqa}
\sup_{z\in K}|\phi^c(z)-\phi^c_n(z)|\leq \langle \eta-\eta_n,g_K\rangle=\int_{X\setminus \Lambda_n}g_K\,d\eta\to 0 ~~\mbox{ for }n\to\infty .
\end{equation} 
This concludes the proof.\kasten

It should be noted that only the stochastic upper bound $\mu_\lambda^f\preceq_v\mu_{\rho \lambda}$ entered in the proof of Theorem \ref{6.1theo}. Analogous constructions
can therefore be carried out with all point random fields which fulfill such a bound.

\

\noindent {\em 5.2 Analyticity of Schwinger functions and relativistic invariance}

\noindent Let $\nu_\lambda^f$ be the modification described in Theorem \ref{6.1theo}. We define the Schwinger (moment) functions of our model:
\begin{equation}
\label{6.3eqa}
S_n(x_1,\ldots,x_n)=\int_{\cal H}\phi(x_1)\cdots \phi(x_n)\, d\nu_\lambda^f(\phi)~,~~x_1,\ldots,x_n\in X.
\end{equation}
That these functions $S_n:X^n\to[0,\infty)$ are well defined is a special case of what is being shown in the following theorem:

\begin{theorem}
\label{6.2theo}
The Schwinger functions $S_n$ are restrictions of holomorphic functions $S_n^c:(X^c)^n\to\C$ to $X^n$.
\end{theorem}
\noindent {\bf Proof.} The natural {\em ansatz} for the analytic continuation of (\ref{6.3eqa}) is
\begin{equation}
\label{6.4eqa}
S_n^c(z_1,\ldots,z_n)=\int_{\cal H}\phi^c(z_1)\cdots\phi^c(z_n)\, d\nu_\lambda^f(\phi)~,~~z_1,\ldots,z_n\in X^c.
\end{equation}
Here, for $\phi\in{\cal H}$,  $\phi^c$ is the analytic extension of $\phi$ from $X$ to $X^c$, cf. the proof of Theorem \ref{6.1theo}. The expression
in (\ref{6.4eqa}) is well-defined: For
$z_1,\ldots z_n\in K$, $K$ a compact in $X^c$, 
\begin{eqnarray}
\label{6.5eqa}
\int_{\cal H}\prod_{j=1}^n|\phi^c(z_j)|\, d\nu_\lambda^f(\phi)&\leq& \int_\Gamma \langle\eta,g_K\rangle^n \, d\mu_\lambda^f(\eta)\nonumber\\
&\leq&\int_\Gamma \langle\eta,g_K\rangle^n \, d\mu_{\rho\lambda}(\eta)\nonumber\\
&\leq& n! \int_\Gamma e^{\langle\eta,g_k\rangle}\, d\mu_{\rho\lambda}(\eta)\nonumber\\
&=&e^{\rho \int_X \exp\{g_K\}-1\, d\lambda}\leq e^{\rho R\int_Xg_Kd\lambda}<\infty,
\end{eqnarray}
by vague stochastic dominance and monotone convergence, see Corollary \ref{3.2cor} and the proof of Theorem \ref{6.1theo}. $R=\|g_K\|_\infty e^{\|g_K\|_\infty}$ with $\|.\|_\infty$ the supremum norm. 

It remains to prove holomorphy of the functions $S_n^c$. Firstly, one can rewrite  them as $S_n^c(z_1,\ldots,z_n)=\int_\Gamma G^c*\eta(z_1)\cdots G^c*\eta(z_n)\, d\mu_\lambda^f(\eta)$ where we set $G^c*\eta=0$ for $\eta \not\in \Pi$. Let us now choose $K\subseteq X^c$ compact
such that $K^n$ contains an open neighborhood of $(z_1,\ldots,z_n)\in (X^c)^n$. Using $|\prod_{j=1}^n G*\eta(z_j)|\leq \langle \eta,g_K\rangle^n$ in connection with 
$\langle\eta,g_K\rangle^n\in L^1(\Gamma,\mu_\lambda^f)$, cf. (\ref{6.5eqa}), we get that the assumptions of Theorem \ref{2.5theo} are fulfilled. The application of this theorem 
now proves holomorphy.
\kasten

Let ${\cal G}$ be a real lie group acting on $X$ such that $\lambda$ is invariant under this action. Denoting the complexification of ${\cal G}$ with ${\cal G}^c$ we assume that 
${\cal G}^c$ is acting holomorphically on $X^c$. Let $Y\subseteq X^c$ be another real submanifold of $X^c$ and ${\cal G}^r=\{\alpha\in{\cal G}^c:\alpha(Y)\subseteq Y\}$ the set-stabilizer
of $Y$ in ${\cal G}^c$. The $\tau$-functions $\tau_n$ are by definitions the restriction of $S_n^c$ to $Y^n$. If the following assumption on $G^c$ holds,
\begin{itemize}
\item[(IV)] $G^c$ is invariant under the real Lie group ${\cal G}$, i.e. $G^c(\alpha(z),\alpha(x))=G^c(z,x)$ $\forall \alpha\in{\cal G},z\in X^c,x\in X$,
\end{itemize}
we get:
\begin{theorem}
\label{6.3theo}
(i) $S_n^c$ is invariant under ${\cal G}^c$, i.e. 
\begin{equation}
\label{6.6eqa}
S_n^c(z_1,\ldots,z_n)=S_n^c(\alpha(z_1),\ldots,\alpha(z_n))~~~\forall \alpha\in{\cal G}^c,z_1,\ldots,z_n\in X^c.
\end{equation}
\noindent (ii) In particular, $\tau_n:Y^n\to\C$ is invariant under ${\cal G}^r$.
\end{theorem}
\noindent {\bf Proof.} Let $z_1,\ldots,z_n\in X^c$ be fixed. Then, $g:{\cal G}^c\ni\alpha\to g(\alpha)= S_n^c(\alpha(z_1),\ldots,\linebreak \alpha(z_n))\in\C$ is 
a holomorphic function on ${\cal G}^c$ as a composition of holomorphic functions. If one can show that $g|_{\cal G}$ is constant,
then Theorem \ref{2.6theo} gives the desired result. For $\alpha\in{\cal G}$ we get
\begin{eqnarray}
\label{6.7eqa}
g(\alpha)&=&\int_\Gamma G^c*\eta(\alpha(z_1))\cdots G^c*\eta(\alpha(z_n))\,d\mu_\lambda^f(\eta)\nonumber\\
&=&\int_\Gamma G^c*(T_\alpha\eta)(z_1)\cdots G^c*(T_\alpha\eta)(z_n)\, d\mu_\lambda^f(\eta)\nonumber\\
&=&\int_\Gamma G^c*\eta(z_1)\cdots G^c*\eta(z_n)\, d\mu_\lambda^f(\eta)=g({\bf 1}),
\end{eqnarray}
where in the second step (IV) has been used, whereas in the third step we applied Corollary \ref{3.1cor}. ${\bf 1}$ denotes the unit element of ${\cal G}$. Hence the assumptions of Theorem \ref{2.6theo}
hold.
\kasten

Obviously, $\tau_n$ is symmetric, i.e. $\tau_n(y_1,\ldots,y_n)=\tau_n(y_{\pi_1},\ldots,y_{\pi_n})$ for any $n$-permutation $\pi$. 

\

\noindent {\em 5.3 Examples}
 \begin{table}
\begin{center}
\begin{tabular}{||l||l|l||}
\hline\hline
Case& Minkowski space&de Sitter space \\
\hline
 $X$&  $\R^d$ &${\mathbb S}^d$ $d$-dim. sphere\\
 \hline
 $\lambda$& Lebesgue  measure& Haar measure\\
 \hline
$X^c$& $\C^d$&$\{ z\in\C^{d+1}:z\cdot z=R^2\}$\\
\hline
$Y$&$\R^d\cong i\R\times\R^{d-1}$ &$\{y\in i\R\times\R^{d}:y\cdot y=R^2\}$\\
\hline
${\cal G}$&$E(d)=O(d)\odot\R^d$& $O(d+1)$\\
\hline
${\cal G}^c$&$L_c(d)\odot\C^d$&$L_c(d+1)$\\
\hline
${\cal G}^r$& $P(d)=L(d)\odot \R^d$&$L(d+1)$\\
\hline
$G^c$&e.g. $\exp\{-(z-x)^2\}$ &e.g. $\exp\{z\cdot x\}$\\
\hline\hline
\end{tabular}
\end{center}
\caption{Simple Examples.  }
\end{table}

\noindent Table 1\footnote{Explanation of symbols: $R>0$ real, $\cdot$ complex analytic extension of the Euclidean scalar product $a\cdot b=\sum_{\kappa=0}^d a^\kappa b^\kappa$, $a,b\in\C^{d+1}$, 
$\odot$ semi-direct product, $E(d)$ Euclidean group on $\R^d$, $L(d)$ the Lorentz group on $\R^d$, $L_c(d)$ its complexification, $(z-x)^2=(z-x)\cdot (z-x)$ where the scalar product is as above but in $\C^d$.} gives possible choices for $G^c$ for the case of Minkowski and de Sitter space-times such that all assumptions (I)-(IV) obviously hold. It is clear that 
(II) is trivial for the case of de Sitter space-time, as the Euclidean space-time is ${\mathbb S}^d$ is compact. 
It is also not very complicated to generate further examples.

In the case of Minkowski space-time one can also chose kernels $G$ which are closer to the requirements of field theory \cite{AGW,AGY1,AGY2}: Let $G_{1/2}(x)$ be the Green's function of
the pseudo differential operator $(-\Delta+m^2)^{1/2}$, $\Delta$ the Laplacian on $\R^d$. $G_{1/2}$ is the standard choice which in the non-interacting case reproduces leads to a Poisson field
measure with the same covariance as the Euclidean free field.
Let furthermore $g_\epsilon(x)=(2\pi \epsilon)^{-d/2}\exp\{- x^2/(2\epsilon)\}$. Then $G_\epsilon=g_\epsilon*G_{1/2}$, with $*$ the usual convolution on $\R^d$,
gives an approximation of $G_{1/2}$ for $\epsilon>0$ small. The following Lemma shows that
such UV-regularized kernels fit into our scheme:

\begin{lemma}
\label{6.1lem}
The kernels $G_\epsilon$ have holomorphic extension $G^c_\epsilon$ to $\C^n$ such that $G^c(z,x)=G^c_\epsilon(z-x)$, $x\in \R^d, z\in\C^d$  fulfills the requirements (I)--(IV). 
\end{lemma} 
\noindent {\bf Proof.}
 (I) Follows from $\sup_{x\in{\R^d}}\|G(.,x)\|_{L^1(\R^d,dx)}=\|G_{1/2}\|_{L^1(\R^d,dx)}\|g_\epsilon\|_{L^1(X,dx)}$. 
 
 To check (II), define $G^c(z,x)=\int_{\R^d}g_\epsilon(z-x-x')G_{1/2}(x')dx'$, $x\in\R^d,z\in\C^d$. For $K\subseteq \C^d$ compact, let
 $g_K(x)$ be defined as in Condition (II). We have $g_K(x)\leq\int_{\R^d}\sup_{z\in K}|g_\epsilon(z-x-x')|G_{1/2}(x')dx'$. Obviously, $ \sup_{z\in K}|g_\epsilon(z-x)|\in L^1(\R^d,dx)$ and therefore the r.h.s. of this inequality is in
 $L^1(\R^d,dx)$ as a convolution of Lebesgue integrable functions. Thus also $g_K\in L^1(\R^d,dx)$. Clearly, $g_K$ is continuous and $g_K(x)\leq \sup_{z\in K+\R^d}|g(z)|\|G_{1/2}\|_{L^1(\R^d,dx)}<\infty$ $\forall x\in X$, hence $g_K\in C_b(X)$.
 
 (III): Taking into account the definition of $G^c(z,x)$ and the fact that $g_{\epsilon}(z)$ is holomorphic on $\C^d$ in combination with the estimate in (II), one gets that
 the assumptions of Morera's theorem \ref{2.5theo} are fulfilled. Holomorphy of $G^c$ in the first argument follows.

 (IV) Let $\alpha\in O(d)$ and $a\in\R^d$. Then 
 \begin{eqnarray}
\label{6.8eqa}
 G^c(\alpha(z)+a,\alpha(x)+a)&=&\int_{\R^d} g_\epsilon(\alpha(z)-\alpha(x)-x')G_{1/2}(x')dx'\nonumber\\
 &=&\int_{\R^d}g_\epsilon (z-x-\alpha^{-1}(x'))G_{1/2}(x')dx'=G^c(z,x)\nonumber\\
 \end{eqnarray} 
here we used the invariance of $g_\epsilon$ under $O(d)$ in the second step and the invariance of the Lebesque measure and of $G_{1/2}$ under rotations and reflections in the third step.
\kasten

Though beyond the scope of this article, it is a natural question to study the $\epsilon\searrow 0$ limit of the analytic continuation of moment functions. E.g. in the Poisson case analytic 
continuations in the "axiomatic" domain \cite{SW} are known to exist \cite{AGW} for $\epsilon=0$. 
  
\

\noindent {\em 5.4 The Wick rotated measure -- failure of invariance}

\noindent The Wick-rotated measure of $\nu_\lambda^f$ can be defined in the following way: Consider the complex valued random field $\varphi=\phi^c|_Y$ over the relativistic space-time $Y$ defined on the probability space
$({\cal H},{\cal B}({\cal H}),\nu_\lambda^f)$ where ${\cal B}({\cal H})$ is the trace sigma algebra of ${\cal B}({\cal M})$ on ${\cal H}$ and $\nu_\lambda^f$ the modification of Theorem \ref{6.1theo}.

By Minlos' theorem, see e.g. \cite{It}, one can associate a Wick-rotated functional measure $\nu_\lambda^{f,\rm Wick}$ e.g. on the measurable space $({\cal D}'(Y,\R^2),{\cal B}({\cal D}'))$, with the random field $\varphi$ where we identified
$\C\cong \R^2$. ${\cal D}'(Y,\R^2)$ is the space of distributions and ${\cal B}({\cal D}')$ the Borel sigma algebra. If one proceeds hastily, one could conclude from Section 5.2 that $\nu_\lambda^{f,\rm Wick}$ is invariant under the Poincar\'e group. This is however wrong in the general case.

To characterize the measure $\nu_\lambda^{f,\rm Wick}$ fully, one has to take into account not only moments of the field $\varphi$ where the multiplication is $\C$-multiplication -- in other words the relativistically invariant functions $\tau_n$ --
but one has to consider mixed expectation values of random fields $\varphi$ and $\bar \varphi$ where the bar denotes complex conjugation. 

It is clear from Theorem \ref{6.1theo} that $\phi$ is also the restriction of the antiholomorphic extension $\bar \phi^c$ defined on $X^c$ to $X$. Defining the complex valued function 
\begin{equation}
\label{6.9eqa}
Q_{n,k}^c(z_1,\ldots,z_n)=\int_{\cal H}\bar \phi^c(z_1)\cdots \bar \phi^c(z_k)\phi^c(z_{k+1})\cdots\phi^c(z_n) \, d\nu_{\lambda}^f(\phi)~,~~z_1,\ldots,z_n\in X^c
\end{equation}
$Q_{n,k}^c:(X^c)^n\to \C$,
one can equally obtain expectation values
$\int_{{\cal D}'}\bar \varphi(y_1)\cdots\bar \varphi(y_k)\linebreak \times\varphi(y_{k+1})\cdots \varphi(y_n)\, d\nu_\lambda^{f,\rm Wick}(\varphi)$ by restriction of $Q_{n,k}^c$ to $Y^n$.

$Q_{n,k}^c$ is anti-holomorphic in the first $k$ and holomorphic in the last $n-k$ arguments. This implies that
${\cal G}^c\ni\alpha\to g(\alpha)=Q_{n,k}^c(\alpha(z_1),\cdots,\alpha(z_n))$ is (anti-) holomorphic in $\alpha$ only if $k=0$ ($k=n$). Hence invariance under ${\cal G}$ does not extend to ${\cal G}^c$ if $k\not\in\{0,n\}$.

Let us illustrate this point further for the examples on Minkowski space given in Section 5.3: One has for $\phi\in{\cal H}$ that $\bar \phi^c(z)=\phi^c(\bar z)$ and hence
$\bar \varphi(y)=\varphi(\theta y)$ with $\theta$ the time reflection on $Y=i\R\times\R^{d-1}\cong \R^d$, $\theta (y^0,\vec y)=(-y^0,\vec y)$, $y=(y^0,\vec y)\in \R^d$. For a Lorentz boost $\alpha$ one obtaines
\begin{eqnarray}
\label{6.10eqa}
&&\int_{{\cal D}'}\bar \varphi(\alpha(y_1))\cdots\bar \varphi(\alpha(y_k))\varphi(\alpha(y_{k+1}))\cdots\varphi(\alpha(y_n)) \, d\nu_{\lambda}^{f,\rm Wick}(\varphi)\nonumber\\
&&~~~~~~~~~~=\int_{{\cal D}'} \varphi(\theta\alpha(y_1))\cdots\varphi(\theta \alpha(y_k))\varphi(\alpha(y_{k+1}))\cdots\varphi(\alpha(y_n)) \, d\nu_{\lambda}^{f,\rm Wick}(\varphi)\nonumber\\
&&~~~~~~~~~~=\int_{{\cal D}'}\varphi(\theta_\alpha y_1 )\cdots \varphi(\theta_\alpha y_k)\varphi(y_{k+1})\cdots\varphi(y_n) \, d\nu_{\lambda}^{f,\rm Wick}(\varphi)\nonumber\\
&&~~~~~~~~~~=\int_{{\cal D}'}\bar \varphi(\theta\theta_\alpha y_1)\cdots\bar \varphi(\theta\theta_\alpha y_k)\varphi(y_{k+1})\cdots\varphi(y_n) \, d\nu_{\lambda}^{f,\rm Wick}(\varphi)
\end{eqnarray} 
with $\theta_\alpha=\alpha^{-1}\theta\alpha$. As the time reflection does not commute with boosts and thus  $\theta\theta_\alpha\not={\bf 1}$, the right hand side of Equation (\ref{6.10eqa}) in general does not coincide with $\int_{{\cal D}'}\bar \varphi( y_1)\cdots\bar \varphi(y_k)\varphi(y_{k+1})\cdots\varphi(y_n) \, d\nu_{\lambda}^{f,\rm Wick}(\varphi)$.  This shows how the Wick rotated measure in general 
fails to be Lorentz invariant. 

The Wick rotation suggested in this note therefore works reasonably only in the case of moment (Schwinger) functions. It does not give a recipe for the construction
of relativistically invariant random fields.

\

\small
\noindent {\bf Acknowledgments.} 
This paper owes a number of improvements to comments of H. O. Georgii on Section 3. Discussions with S. Albeverio, T. Kuna, H. Thaler and M. W. Yoshida are gratefully acknowledged.
This work has been made possible through financial support of D.F.G through projects "Stochastic analysis and systems of infinitely many degrees of freedom", "Stochastic methods in QFT" and SFB 611 A4. 
It is my pleasure to thank the organizers 
for the opportunity to report parts of this work at the 1st Sino-German Meeting on Stochastic Analysis.

\end{document}